\documentclass[12pt]{article}
\usepackage[utf8]{inputenc}
\usepackage{setspace}
\usepackage{palatino}
\usepackage{graphicx}
\usepackage{float}
\usepackage{titling} 
\usepackage{multirow}
\usepackage{lscape}
\usepackage{amsmath}
\usepackage{accents}
\usepackage{amssymb}
\usepackage{fancybox}
\usepackage{algorithm}
\usepackage{amsthm}
\usepackage{subcaption}
\usepackage[dvipsnames]{xcolor}
\usepackage{bbm}
\usepackage{natbib}
\usepackage{algpseudocode}
\usepackage{subcaption}
\usepackage{booktabs}
\usepackage[left=0.8in,top=1in,right=0.8in,bottom=1in]{geometry}
\fontfamily{ppl}\selectfont 
\usepackage[colorlinks,linkcolor=blue,citecolor=blue,urlcolor=blue]{hyperref}

\newcommand{\Prob}[1]{\Pr\left(#1 \right)}

\newcommand{\Expect}[1]{{\rm I\kern-.3em E}\left(#1\right)}
\newcommand{\comment}[1]{}

\usepackage{setspace}

\DeclareMathOperator*{\argmax}{arg\,max}

\title{Where to place a mosquito trap for West Nile Virus surveillance? \vspace{-5mm}}

\author{Anwesha Chakravarti, Bo Li, Dan Bartlett, Patrick Irwin, Rebecca Smith}

\date{\vspace{-5ex}}

\begin{document}
\setstretch{1.25}
\maketitle
\begin{abstract}
The rapid spread of West Nile Virus (WNV) is a growing concern. With no vaccines or specific medications available, prevention through mosquito control is the only solution to curb the spread. Mosquito traps, used to detect viral presence in mosquito populations, are essential tools for WNV surveillance. But how do we decide where to place a mosquito trap? And what makes a good trap location, anyway?

We present a robust statistical approach to determine a mosquito trap's ability to predict human WNV cases in the Chicago metropolitan area and its suburbs. We then use this value to detect the landscape, demographic, and socioeconomic factors associated with a mosquito trap's predictive ability. This approach enables resource-limited mosquito control programs to identify better trap locations while reducing trap numbers to increase trap-based surveillance efficiency. The approach can also be applied to a wide range of different environmental surveillance programs.
\bigskip

\noindent {\large\bf Keywords}

\medskip

\noindent West Nile Virus, Spatio-temporal analysis, Causal inference, Surveillance. 
\end{abstract}

\section{Introduction}\label{sec:Introduction}
    Since its US emergence in the New York Metropolitan area in the fall of 1999, West Nile Virus (WNV) has spread rapidly. With over 52,000 cases nationwide to date, including approximately 2,700 in Illinois alone, WNV stands as the leading cause of mosquito-borne disease in the continental United States \citep{WNV}. Thus, taking adequate measures for the control of WNV has become vital. The primary mode of transmission for this virus is through host-seeking female \textit{Culex} spp. mosquitoes. Cases in Illinois are mainly seen from late summer to early fall, as this is the peak of both mosquito season and WNV infection within mosquitoes and the primary avian hosts \citep{WNV_ill}. Most people infected by WNV show no symptoms, which results in under-reporting of WNV infection and limited information about the true extent of the spread of the infection in humans \citep{zou2010west,petersen2013west}. The absence of vaccines or medications to treat the virus makes prevention through mosquito control the only solution for controlling the spread of the infection.

     To monitor mosquito populations for WNV, Mosquito Abatement Districts (MADs) and Public Health Departments deploy mosquito traps across different spatial extents throughout their jurisdictions.
     The use of mosquito traps is vital for monitoring the spread of the virus and providing information for response efforts \citep{williams2007comparison,Traps}. The traps capture a variety of mosquito species, from which the mosquitoes of the \textit{Culex} spp. are identified and collected in pools of up to 50. These pools are then tested for the presence of the virus. The resulting data from trap testing is used to predict the risk of a human case in the vicinity of the location of the trap, which is then used to inform mosquito abatement practices. Although there is extensive research focused on modelling the incidence of human cases of WNV using trap-related data, along with climate, socio-economical and geographical variables \citep{ruiz2004environmental,keyel2021proposed,uelmen2021effects}, and identifying hotspots for WNV \citep{curtis2014comparison,krebs2014host,hamer2014dispersal}, there has been little research on identifying the most effective locations to place the traps. Given the significance of mosquito traps in the WNV surveillance process, strategic placement of traps can greatly aid in disease control and prevention.

    The management of each mosquito trap is expensive in labor. Currently, mosquito traps are placed using a grid-based approach. However, in areas where trap deployment has to be minimal due to a lack of resources, this grid approach may result in poor surveillance accuracy and thus limit the predictive performance for WNV cases. Given the ongoing decline in funding for vector-borne disease surveillance, management, and research \citep{hadler2014national,dye2022riding}, enhancing the cost-effectiveness of each trap is crucial. If we can identify optimal locations for trap placement, health departments will be able to streamline their surveillance efforts and minimize the number of traps required. Therefore, in this work, we aim to identify a statistical approach to determine optimal trap placement - a line of research that has not been systematically conducted in the field of West Nile Virus mosquito modelling. 

    Numerous factors could influence a trap's ability to provide accurate predictions of WNV cases, such as the terrain in which it is situated, the local mosquito and bird populations, and the demographic characteristics of the human population, among others. To determine the best locations to place the traps, we investigate the factors significantly affecting trap performance through a three-phase approach. In the initial two phases, we assess a trap's ability to forecast the occurrence of a human case of WNV and assign a score to each trap. This involves utilizing a spatiotemporal model to make predictions of human WNV occurrence based on trap data, with the score defined as a weighted mean of sensitivity and specificity values. Subsequently, in the third phase, we utilize these scores to identify variables influencing the trap's overall performance via a causal model. 

    Besides the study of mosquito traps, another of our primary contributions is the modelling framework itself, which can be tailored to suit other specific needs. 
    Each phase of the process, including the definition of the score and the variables we explored to identify the best trap locations, can be customized based on the requirements of mosquito control departments. Therefore, although we have specifically applied our method to traps within the Chicago metropolitan area and its suburbs, our approach can be generalized to analyse any geographical region of interest worldwide. Furthermore, the framework we have developed has the potential to address similar challenges in various other domains, particularly in the area of environmental surveillance. 
    
    Our paper is structured as follows: Section 2 describes the dataset and gives information on the data collection process from the traps. Section 3 describes the modelling framework explaining each of the phases in detail. In Section 4, we present the results of each phase and offer insights derived from these results to understand the data better and to achieve our objective of finding the best trap locations. Section 5 explores an additional application of our modelling framework that can help inform strategies for improving the effectiveness and accuracy of WNV surveillance efforts in the region.  Through this extended application, we showcase the versatility and robustness of our modelling approach in addressing complex real-world challenges beyond its initial scope. We conclude with a discussion in Section 6. 

\section{Dataset}\label{sec:Dataset}

\begin{figure}[t]
    \centering 
    \vspace{-1.6cm}
    \includegraphics[scale = 0.9]{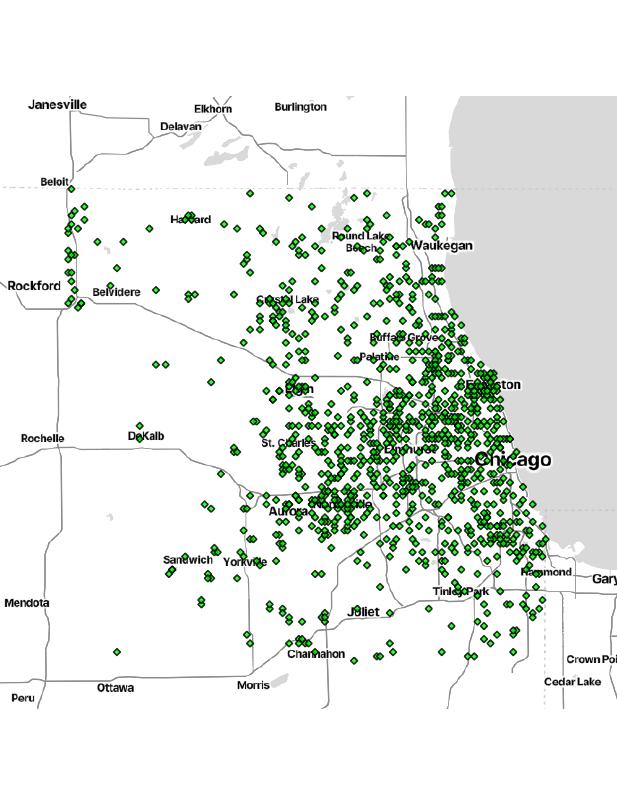}\vspace{-1.6cm}
    \caption{Location of the traps in Chicago, Illinois and its suburbs.}
    \label{fig:loc}
\end{figure} 

Our dataset spans from 2004 to 2018 and comprises data collected from traps placed in Chicago and its surrounding suburbs (Figure \ref{fig:loc}). For our analysis, we consider a total of 1062 traps distributed across 275 distinct zip codes, with Cook and DuPage counties having the highest trap density. The traps are set weekly, and mosquitoes are collected from these traps every 1-2 trap nights and then sorted to identify the \textit{Culex} spp. Subsequently, the \textit{Culex} mosquitoes are pooled into groups of up to 50 and undergo testing for the presence of WNV for each pool \citep{CDCMosqtest}\footnote{Further details can be found at \url{https://www.cdc.gov/mosquitoes/guidelines/west-nile/laboratory-diagnosis-testing/non-human-laboratory-diagnosis.html}.}. Some traps may have multiple pools of mosquitoes tested from the same collection trap-night, if more than 50 \textit{Culex} mosquitoes were collected on that trap-night. This is because WNV testing is considered not sufficiently sensitive for pools of more than 50 mosquitoes \citep{sutherland2007detection}. Since the WNV test is carried out at the pool level rather than for each mosquito, each positive pool only indicates that there is at least one infected mosquito in the pool. The mosquito infection prevalence for a given trap-night based on a pooled sample is calculated using the standardized maximum likelihood estimate (MLE) approach  \citep{nasci2005calculation}.
Data available for analysis in this system include:
\begin{itemize}
    \item \textbf{Mosquito and WNV data} containing information on the mosquito pools, the test results for each of these pools and whether a human WNV case was detected in the vicinity of the traps.
    \item \textbf{Landcover data} containing data on the landcover characteristics of the trap sites. It includes variables like canopy cover, percentage of impervious land, and indicators for whether the traps were placed in highly developed areas, near open water bodies, etc. 
    \item \textbf{Demographic data} containing information on the total population, population by race, age and so on, in a 1500m buffer area around the trap.
    \item \textbf{Socio-economic data} containing data on poverty levels, percentage occupied housing and education levels of the population in a 1500m buffer area around the trap.
\end{itemize}

The mosquito and WNV data were obtained through a data-sharing agreement with the Illinois Department of Public Health. The landcover data was collected from the National Land Cover Dataset (NLCD)\footnote{Data available at \url{https://www.mrlc.gov/data/nlcd-2019-land-cover-conus}.} \citep{dewitz2019national}, the census block-level population data was retrieved from the \citet{Popndata}\footnote{Data available at \url{https://www.census.gov/programs-surveys/decennial-census/about/rdo/summary-files.html}.}, and  
the census data for educational attainment and poverty was collected from the \citet{Eddata}\footnote{Data available at \url{https://www.census.gov/newsroom/press-kits/2020/acs-5-year.html}.}.

\section{Modelling Methodology}\label{sec:Modelling Methodology}
A major challenge in determining the optimal trap locations is quantifying the ``goodness" of a trap's placement. Our proposed method addresses this problem by assigning a score to each trap based on the trap's historical performance in predicting human cases of WNV, followed by utilizing this score to determine the factors that cause variability in a trap's prediction performance.
Data associated with high-score locations will be used to specify the characteristics of a good trap location. To achieve this, we present a three-phase approach:

    \begin{itemize}
        \item \textbf{Phase 1:} Model historic performance of the traps for predicting human cases.
        \item \textbf{Phase 2:} Assign score for individual traps.
        \item \textbf{Phase 3:} Characterize locations that correspond to higher score traps.
    \end{itemize}

\subsection{Phase 1: Modelling the historic performance of the traps}\label{sec:Phase 1}

We construct a logistic model to quantify the historical performance of the traps in predicting the incidence of human cases of WNV in the proximity of the traps. The explanatory variables of primary interest in the model are the mosquito pool data, which can be either single or multiple, depending on how many pools for a trap were tested. We also include other explanatory variables to control the effect due to the variation of pool size, the week and year of the data collection, and the location of traps. 
The response variable is a binary variable indicating whether a human case was detected within a 1.5 km radius around the trap within two weeks of the pool's collection. For instance, if a pool was collected during week 31, the response would be the incidence of a human case during weeks 32 and 33. The two-week leading time was chosen due to the biology of mosquito-borne spreading of WNV and the latency period of human WNV infections \citep{karki2020drivers}. Specifically, the variables used in the model are: 
\begin{itemize}
    \item \textbf{Pool Size} - Number of mosquitoes in the pool.
    \item \textbf{Test Indicator} - Indicator variable denoting whether the pool tested positive for the presence of WNV. 
    
    \item \textbf{Risk} - The risk associated with a pool. Since individual mosquitoes do not undergo testing, the prevalence of the virus in the tested pool remains unknown. To measure the risk associated with a pool, we first estimate the prevalence of infected mosquitoes in the pool using the Maximum Likelihood Estimation (MLE) method. MLE is the standard method in WNV surveillance to estimate the prevalence of the virus in a pool. Note that the MLE in this context is different from the MLE in statistics. Details regarding the MLE method and its advantages can be found in \citet{gu2008fundamental}. Given the MLE value, the risk for a pool in mosquito surveillance is defined using the following Vector Index equation \citep{nasci2005calculation}:
 $$\hbox{Risk}= \frac{\hbox{Daily\ Abundance*MLE}}{1000},$$
where$$\hbox{Daily\ Abundance} = \hbox{Avg\ Abundance}\times \  \hbox{Number\ of\ pools\ in\ that\ trap\ on\ that\ day},$$ 
$$\hbox{Avg \ Abundance} = \frac{\hbox{Number \ of\ total\ mosquitoes\ caught \ by\ that\ trap\ in\ that \ week}}{\hbox{Number\ of\ pools\ in\ that\ trap\ in\ that \ week}}.$$
    \item \textbf{Week and Year} - Week in the year, and the year when the pool was collected.
    \item \textbf{Latitude and Longitude} - Location of the trap.
\end{itemize}

\subsubsection{Model description and evaluation} \label{sec:Model description}
We model the data using a Spatial Generalized Linear Mixed-effects Model (Spatial GLMM) with a binomial family \citep{f2007methods,spamm}. In simpler terms, Spatial GLMM is a logistic regression model with spatially correlated errors. Since closely situated traps are likely to exhibit similar behaviour, we expect to observe spatial correlation in our data due to the proximity of the traps. Thus, using a GLMM with spatially autocorrelated random effects is more appropriate. We choose Mat\'ern kernel \citep{genton2001classes,minasny2005matern}, the most popular spatial covariance function, to model the random effects over the trap locations. We take the explanatory variables described above as fixed effects in the model. 
Let $Y(pl)$ be the binary (0/1) response variable with $Y(pl)=1$ indicating that a human case was detected in a 1.5 km radius of the trap within a two-week time frame following the collection of the pool $(pl)$. Let $p(pl):= \Prob{Y(pl) = 1}$. Then, our statistical model is as follows:
\begin{equation*}
     \hbox{logit}\big(p(pl)\big)= \beta_0 + \beta_1\hbox{Poolsize}(pl) + \beta_2\hbox{TestInd}(pl) +\beta_3\hbox{Risk}(pl) + \beta_4\hbox{Week}(pl)+\beta_5\hbox{Year}(pl) + b(pl),
\end{equation*}
where $\hbox{logit}(x) = \log\left(x/(1-x)\right)$ and the random effect $b(pl)$ is modelled by a Gaussian random process
\begin{equation*}
    b(pl) \sim N(0,\Sigma_b), 
\end{equation*}
where $\Sigma_b$ is governed by the Mat\'ern covariance function based on the Euclidean distance between pools. 
That is, the correlation between two pools with distance $d>0$ is computed by 
$$Corr(d) = \frac{d^\nu K_{\nu}(d)}{2^{\nu-1}\Gamma(\nu)},$$
where $K_\nu$ is the modified Bessel function of the second kind of order $\nu$ and $\Gamma(\cdot)$ is the gamma function. 
The above mixed-effect model that includes both fixed and random effects is fitted via a likelihood-based method using the R package ``spaMM" \citep{spamm}.



We measure the performance of the traps using the sensitivity and specificity of the classifier, the logistic model,  at each trap location. Following the 5-fold cross-validation, the data of each year is first randomly divided into training and testing sets in a 4:1 ratio and then all annual training and testing data are pooled together to form the overall training and testing data to be used in our model fitting and evaluation. 
The model is fitted with the training data, and the predictions for the testing data are evaluated to calculate the sensitivity and specificity for a particular trap. This procedure is repeated five times, and the average sensitivity and specificity scores for each trap (denoted by $t$) are computed. Mathematically, this can be described as follows:
\begin{equation*}
    \hbox{AvgSens}_{t} = \frac15\sum_{cv = 1}^5\left(\frac{TP_t}{P_t}\right)_{cv} \quad and \quad \hbox{AvgSpec}_{t} = \frac15\sum_{cv = 1}^5\left(\frac{TN_t}{N_t}\right)_{cv},
\end{equation*}
where $TP_t$ is the number of true positive cases corresponding to trap $t$ that are predicted as positive, $TN_t$ is the number of true negative cases corresponding to trap $t$ that are predicted as negative, and $P_t$ and $N_t$ are the total number of true positive and negative cases, respectively,  for trap $t$. 

\subsubsection{Problem of class imbalance and thresholding}\label{sec:imbalance}

\begin{figure}[t]
    \centering
    \begin{subfigure}{0.48\textwidth}
        \centering
        \includegraphics[width=1\textwidth]{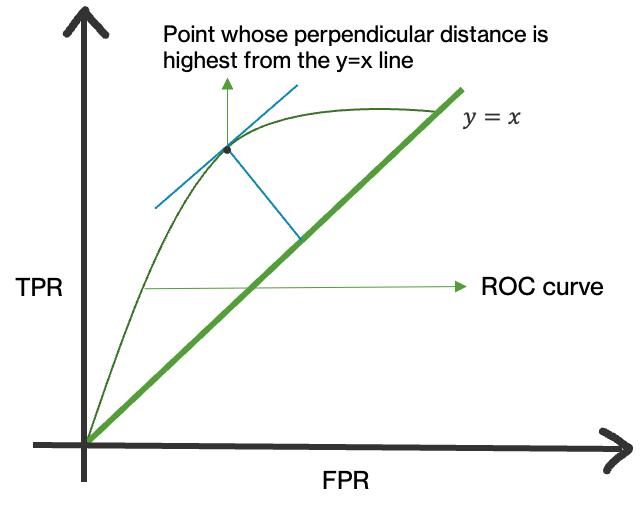} 
        \caption{Sensitivity and specificity having equal importance}
        \label{Fig1}
    \end{subfigure}\hfill
    \begin{subfigure}{0.48\textwidth}
        \centering
        \includegraphics[width=1\textwidth,height = 6.4cm]{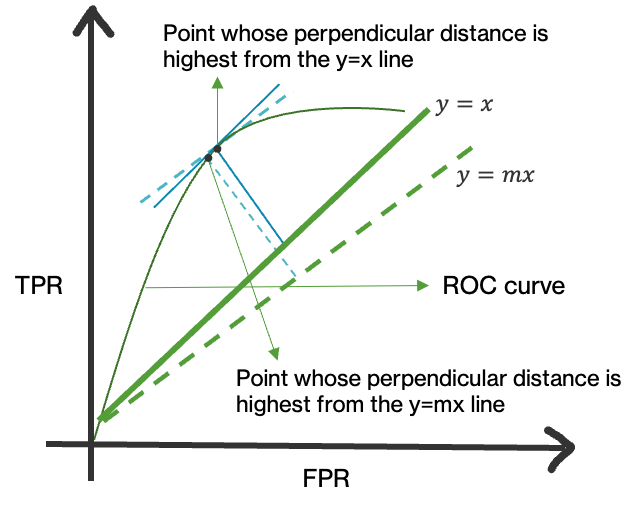} 
        \caption{ \centering Specificity set as $m$ times as important as sensitivity with $m <1$. }
        \label{Fig2}
    \end{subfigure}
    \caption{Choosing the threshold using ROC curve.}
    \label{roc}
\end{figure}

Since the occurrence of human cases of WNV is low, the model fitting suffers from class imbalance. This is a common issue in the field of medicine where the occurrence of a disease is significantly lower than the absence of the disease \citep{sun2009classification}. In particular, we have 184101 instances of $Y(pl)=0$ compared to only 2951 instances of $Y(pl)=1$. Consequently, the probabilities predicted by the logistic regression model tend to be higher for the event $Y(pl)=0$. An effective approach to mitigating this issue is to set an appropriate threshold (instead of the default value 0.5) on the predicted probabilities to obtain the predicted class labels of 0 or 1 \citep{goorbergh2022harm}. Such threshold can be identified using the Receiver operator characteristics (ROC) curve in Figure \ref{roc}. The ROC curve plots the False Positive Rate (FPR) on the $X$–axis vs the True Positive Rate (TPR) on the $Y$-axis at different thresholds. A random classifier model would have the $y = x$ line as its ROC curve. Thus, to obtain the desired threshold, one would choose the point which has the highest perpendicular distance from the $y = x$ line (Figure \ref{Fig1}). 
Let $TPR(k)$ and $FPR(k)$ denote the true positive rate and the false positive rate, respectively, at a threshold value $k$. Then, the optimal threshold can be computed as follows: 
\begin{equation}\label{eq:1}
   \hbox{threshold}_{opt} =  \argmax_{k}\left(TPR(k) - FPR(k)\right),
\end{equation}
which corresponds to the point whose ROC has the highest perpendicular distance from the $y = x$ line. 

Due to the nature of the problem, our priority is to accurately predict the occurrence of human cases, even if it may come at the expense of predicting some false negatives. That is, we want to increase our sensitivity (True Positive Rate) at the expense of reduced specificity (True Negative Rate). To achieve this goal, we modify (\ref{eq:1}) to give more weight to the sensitivity. That is, the optimal threshold is now computed as follows: 
\begin{equation}\label{eq:thres}
    \hbox{threshold}_{opt} = \argmax_{k}\left(TPR(k) - m*FPR(k)\right), 
\end{equation}
where $m<1$ is a parameter that controls the importance of sensitivity as opposed to specificity. The parameter $m$ makes our methodology flexible to the needs of the mosquito control organizations. The value of $m$ can be chosen through a cost analysis of losses incurred due to false positives as opposed to losses due to false negatives. The final prediction labels are then assigned as: $Y_{pred} = 1$ if the predicted $P(Y(pl) = 1) > \hbox{threshold}_{opt}$, and $Y_{pred} = 0$ otherwise. 
This is equivalent to choosing the optimal threshold as the point with the highest perpendicular distance from the $y=mx$ line where $m<1$, since $y = mx$ represents the ROC curve for a random classifier with specificity being $m \times100\%$  as important
as sensitivity (Figure \ref{Fig2}).





\subsection{Phase 2: Assigning a score for each individual trap}
Given the model developed in Phase 1, we use the average sensitivity and specificity values for each trap as defined in Section \ref{sec:Model description} to assign a score to the individual trap $t$:
\begin{equation}\label{eq:score}
    \hbox{score}_t = \frac{ m*\hbox{AvgSpec}_{t} + \hbox{AvgSens}_{t}}{m+1},
\end{equation}
where $m<1$ is as defined in Section \ref{sec:imbalance}. The defined score is a weighted mean of the sensitivity and specificity values of the trap, with the weight $m$ controlling the importance of false negatives as compared to the false positives. In Section \ref{sec:Results Phase 1}, we explain our choice of the score function using our data as an example.

\subsection{Phase 3: Characterizing locations with higher trap scores}\label{sec:Phase 3}

We now proceed to identify the best locations to place traps with respect to the local demographic, socio-economic and landcover variables.
To investigate which factors may cause the variability of the trap scores as calculated in Phase 2, we analyze the causal effect of the variables on the score. Denote the covariates by $X = (X_1,X_2,\dots,X_p)$ and let $x_i$ and $x_i'$ ($x_i \neq x_i'$) be two possible values that the covariate $X_i$  may take. The causal effect of moving covariate $X_i$ from $x_i$ to $x_i'$ is $Y(x_i) - Y(x_i')$ where $Y(x_i)$ is the potential outcome when covariate $X_i$ takes the value $x_i$. 
We estimate the average causal effect for moving the covariate $X_i$ from $x_i$ to $x_i'$ using $$\Expect{Y(x_i)} - \Expect{Y(x_i')}.$$ Since our covariates are continuous variables, it is not feasible to estimate the average causal effect for all such pairs of values. A practical approach \citep{crump1976fundamental,altshuler1981modeling,dominici2002air,wang2015exposure} to computing the average causal effect for moving between any pair of values of a covariate is to calculate the average dose-response function (ADRF, also known as the exposure-response function) defined as \begin{equation}\label{eq:ADRF}
    \mu({x_i}) = \Expect{Y(x_i)}.
\end{equation}
 The ADRF contains information on the average causal effect for all comparisons of $x_i$ since the average causal effect of moving covariate $X_i$ from $x_i$ to $x_i'$ can simply be computed as $\mu(x_i)-\mu(x_i')$.
It is important to note that the ADRF defined in (\ref{eq:ADRF}) is different from the widely used regression function $\mu^*(x_i) = \Expect{Y|X_i = x_i}$, which gives the conditional expectation of the response at a given covariate value. Causal effects estimated using the ADRF involve comparing the potential outcomes when a covariate value is changed within a single population of units. However, regression curves may represent the outcomes for varying populations at different values of the independent variable $X_i$. When the covariates $X$ are independent of the response $Y_i(x)$ at each $x$, the regression curve $\mu^*(x)$ and the ADRF $\mu(x)$ coincide. This is common in randomized experiments but is rarely seen in observational studies like the one under analysis. A discussion on the difference between regression curves and ADRFs can be found in Section 1.2.7 of \citet{galagate2016causal} for added insights. 

\begin{figure}
  \centering
  \includegraphics[width = 0.9\linewidth]{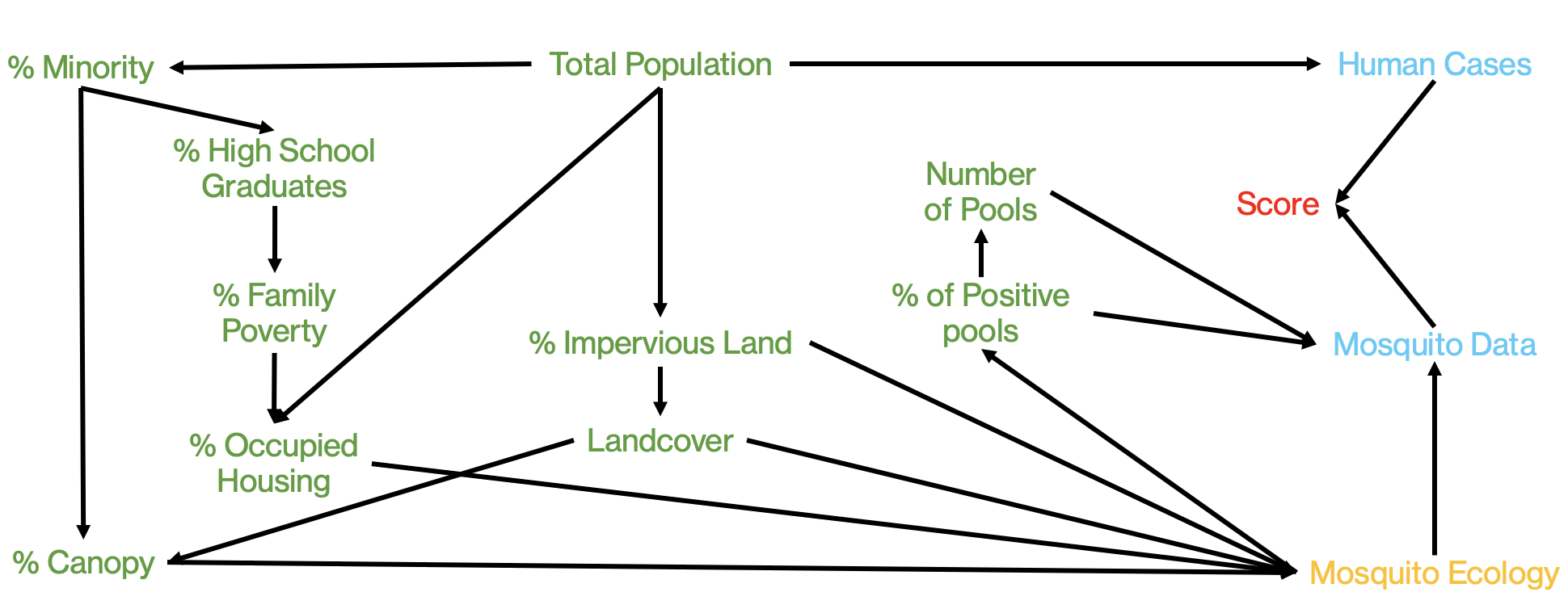}
  \caption{The DAG which represents the causal relationships between our variables. The names in green denote the covariates considered in our model, the names in blue denote the hidden variables which directly affect the score, the names in yellow denote the other hidden variables and the name in red denotes our response.}
  \label{fig:DAG}
\end{figure}

A vital part of causal inference is identifying the confounding variables which could affect the estimate of the causal effect of a covariate on the score \citep{greenland1999confounding,pearl2000models,vanderweele2013definition}. A confounder is a variable which influences both the covariate and the response leading to spurious associations. Hence, they need to be conditioned on, to remove the confounding and accurately estimate the causal effect of the covariate on the response. Consequently, we draw a directed acyclic graph (DAG) (Figure \ref{fig:DAG}) to graphically represent and visualize the causal and possible confounding relationships between the variables in our model. 
\begin{figure}
  \begin{subfigure}{0.48\textwidth}
        \centering
        \includegraphics[scale = 0.32]{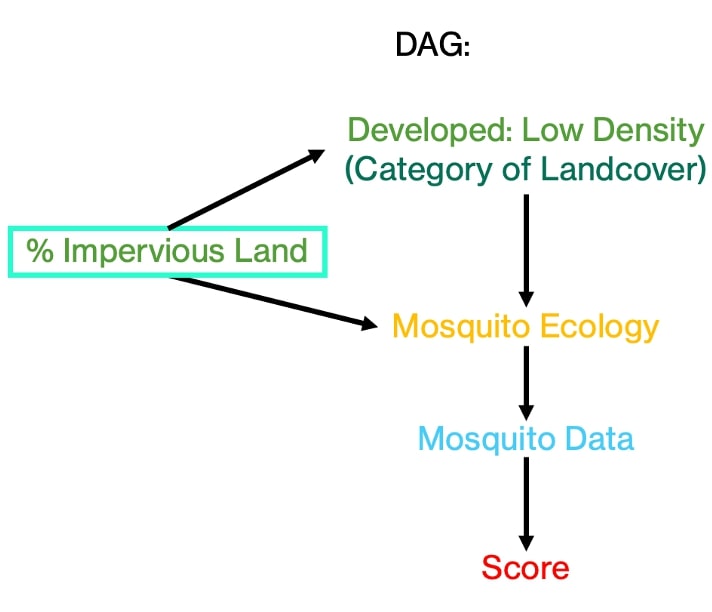} 
        \caption{DAG for the effect of the trap being placed on low-density landcover on score}
        \label{Fig:DAGlowdens}
    \end{subfigure}\hfill
    \begin{subfigure}{0.48\textwidth}
        \centering
        \includegraphics[scale = 0.32]{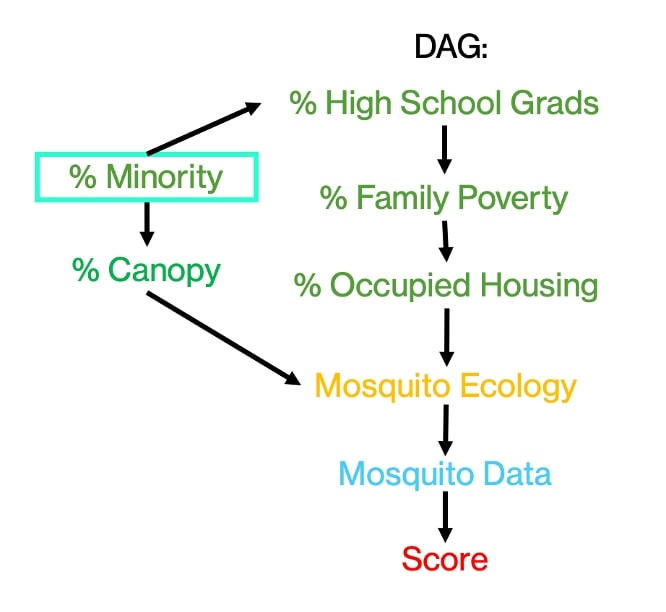} 
        \caption{DAG for the effect of percentage of high school graduates on score}
        \label{Fig:DAGhsgrad}
    \end{subfigure}
      \caption{Subsets of the DAG shown in Figure 3. The teal box indicates the variables being conditioned on to control for confounding.}
  \label{DAG subset}
\end{figure}

To help understand the DAG, in Figure \ref{DAG subset} we present two smaller DAGs which are subsets of the full DAG as examples. These condensed DAGs illustrate the causal pathways from a variable of interest to the score. Figure \ref{Fig:DAGlowdens} illustrates the causal pathway of the effect of a trap being placed in a low-density landcover region on the score. Here, we observe that the variable ``percentage of impervious land" affects both the ``low density" variable and the score, making it a confounder that requires conditioning to obtain accurate estimates for the causal relationship between the ``low density" and the score. The teal box surrounding the confounder indicates that it has been conditioned on. Similarly, in Figure \ref{Fig:DAGhsgrad}, we display the causal pathway for the effect of the percentage of high school graduates on the score. This DAG is slightly more intricate than the first one, featuring two variables, ``percentage of minority population" and ``percentage of canopy cover," both as confounders. However, since there is only one path passing through these variables, conditioning on any one of them effectively blocks the confounding effect, so in this example we only condition on ``percent of minority". The DAG approach allows us to identify potential confounders in estimating the causal effect of different variables on the score.
By analyzing the DAG for each variable, we can identify potential confounders that may influence the estimation of the causal effect of the variables on the score.

We use the R package ``causaldrf" \citep{causaldrf,galagate2016causal} to estimate the ADRF. In particular, we use a doubly robust method \citep{robins2000robust} to estimate the ADRF using a generalized additive model (GAM) estimator. Further details on this method of estimation can be found in \citet{hirano2004propensity,galagate2016causal} and \citet{flores2012estimating}. The ADRF helps us identify the variables that have an impact on the score and thus points us towards a good trap location. 

\section{Results}
\subsection{Sensitivity and specificity values from Phase 1 and trap scores from Phase 2}\label{sec:Results Phase 1}
\begin{figure}
    \centering
    \vspace{-1.5cm}
    \includegraphics[width=0.6\linewidth]{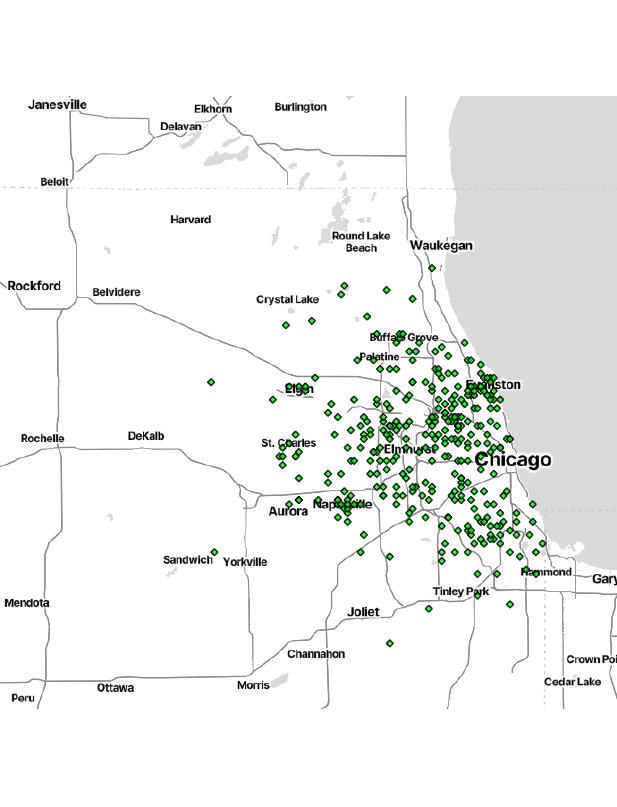}
    \vspace{-1.5cm}
    \caption{The 326 traps $(T^*)$ that have at least one human case within a 1.5 km radius during the 2004-2018 period in Chicago, Illinois, USA}
    \label{fig:t-star}
\end{figure}

We analyze the average sensitivity and specificity values obtained from Phase 1 of our
method (Section \ref{sec:Phase 1}) to illustrate the intuition behind our choice of score.
Our data contains many traps that have not had any human cases of WNV in
their vicinity (in a 1.5 km radius around them), i.e., many traps do not have $Y(pl) = 1$ during the observation period. For such traps, it is impossible to compute their sensitivity values due to the lack of true positives.  It turns out we can only compute sensitivity for 326 traps (see figure \ref{fig:t-star}) out of the total 1062 traps. We denote the set of all 1062 traps as $T$ and the subset of 326 traps which have had at least one human case in their vicinity as $T^*$. While we use the data from the entire traps in $T$ to build our model in Phase 1, we compute the scores for only the traps in $T^*$ in Phase 2. 

\begin{figure}
    \centering
    \begin{subfigure}{0.32\textwidth}
        \centering
        \includegraphics[width=1\textwidth]{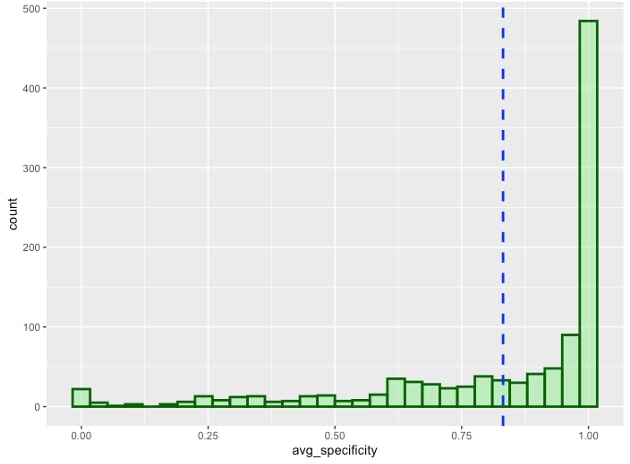} 
        \caption{ \centering Specificity values for the traps $\in T$ (1062 traps)}
        \label{Fig4a}
    \end{subfigure}
    \hfill
    \begin{subfigure}{0.32\textwidth}
        \centering
        \includegraphics[width=1\textwidth]{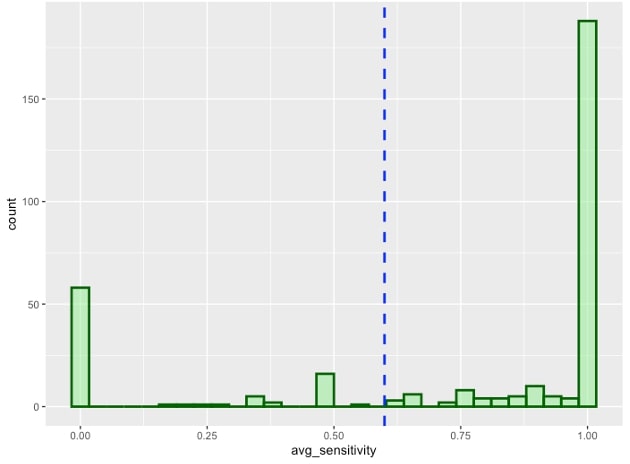} 
        \caption{ \centering Sensitivity values for the traps $\in T^*$ (326 traps)}
        \label{Fig4b}
    \end{subfigure}\hfill
    \begin{subfigure}{0.32\textwidth}
        \centering
        \includegraphics[width=1\textwidth]{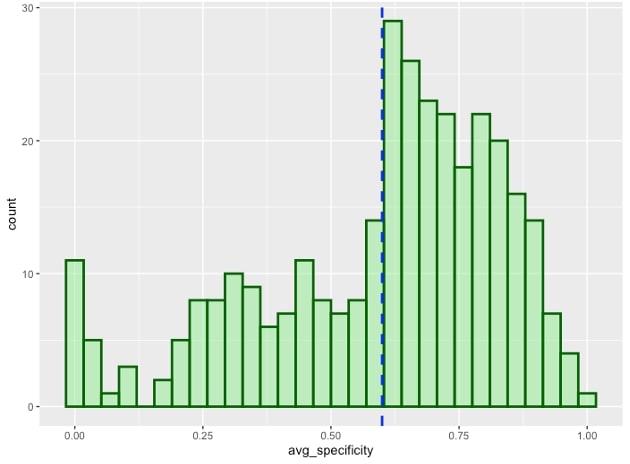} 
        \caption{ \centering Specificity values for the traps $\in T^*$ (326 traps)}
        \label{Fig4c}
    \end{subfigure}
    \caption{Specificity and Sensitivity values computed from the model in Phase 1.}
    \label{sensandspec}
\end{figure}

Figure \ref{sensandspec} reports the sensitivity and specificity scores for different groups of traps. Figure \ref{Fig4a} displays the specificity scores for all traps ($T$). There is a prominent peak at 1, indicating that the majority of traps effectively predict the absence of human cases ($Y(pl)=0$). Figures \ref{Fig4b} and \ref{Fig4c} show results only for the traps in $T^*$. A peak at 1 in the sensitivity values in figure \ref{Fig4b} suggests that most traps in $T^*$ effectively predict the occurrence of human cases ($Y(pl)=1$). Additionally, a smaller peak at 0 in figure \ref{Fig4b} implies that some traps failed to predict the occurrence of any human cases in this time; given the rarity of human cases, many of these extreme values are likely due to traps with only a single human case in their vicinity over the whole time period. However, the specificity values for traps in $T^*$ in figure \ref{Fig4c} are more spread than those in figure \ref{Fig4a}, indicating that traps are better at predicting $Y(pl)=0$ outcomes when they have not encountered a human case ($Y(pl)=1$), a notion that aligns with intuition. This observation motivates our definition of the score in Phase 2 as the weighted average of the specificity and sensitivity values of the traps.

\begin{figure}[t]
    \centering
    \begin{subfigure}{0.48\textwidth}
        \centering
        \includegraphics[width=1\textwidth]{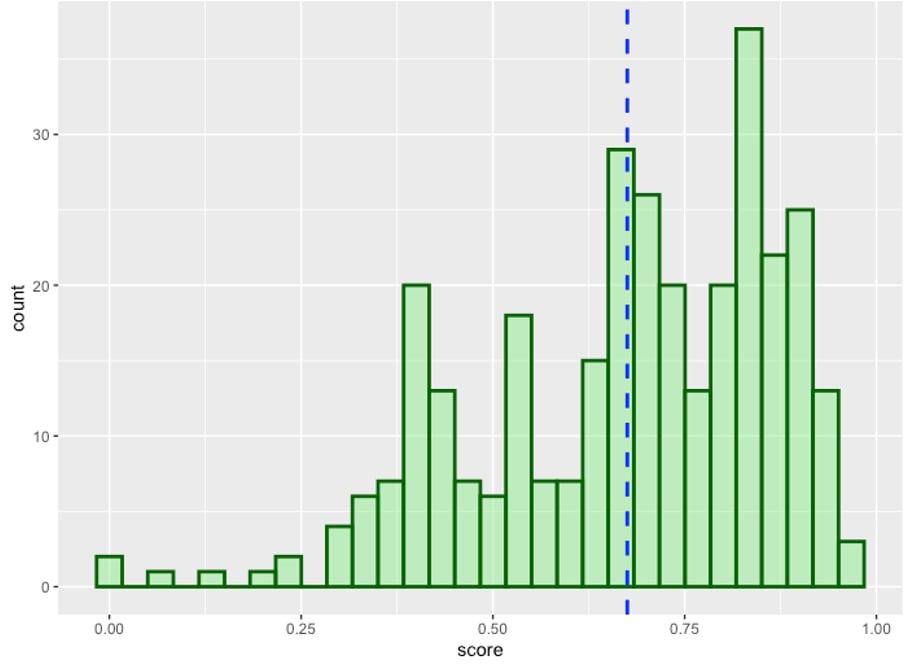}
        \caption{ \centering Histogram of trap scores.}
        \label{Fig5a}
    \end{subfigure}
    \hfill
    \begin{subfigure}{0.48\textwidth}
        \centering
        \includegraphics[width=1\textwidth]{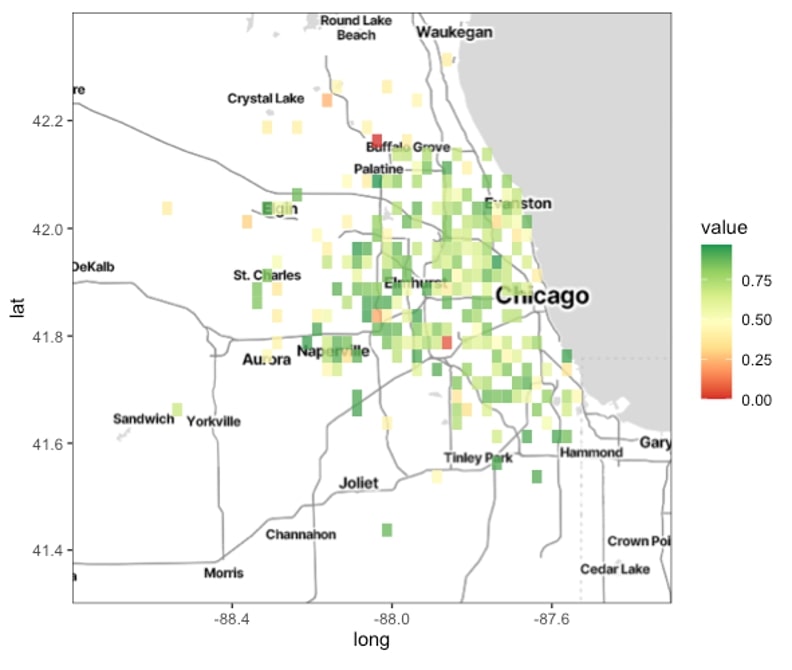} 
        \caption{ \centering Geospatial distribution of trap scores.}
        \label{Fig5b}
    \end{subfigure}
    \caption{Distribution of the scores computed in Phase 2 of traps in $T^*$ (326 traps), assuming  $m = 0.9$ in equation \ref{eq:score}.}
    \label{fig:score}
\end{figure}

The scores plotted in Figure \ref{fig:score} are computed with $m = 0.9$ in equation \ref{eq:score}. The distribution of the scores is left skewed with a mean score of approximately 0.72 (depicted by the blue line). This suggests that while most traps perform well, some traps exhibit poor performance. Figure \ref{Fig5b} shows the spatial map of trap scores, where each marker represents a trap location. The plot shows a clustered pattern of scores, which motivates us to explore which factors result in a regional higher or lower score. This leads to the Phase 3 of our method. 

\subsection{Identifying efficient trap locations using Phase 3}

\begin{figure}[ht]
    \centering
    \begin{subfigure}{0.32\textwidth}
        \centering
    \includegraphics[width=\linewidth]{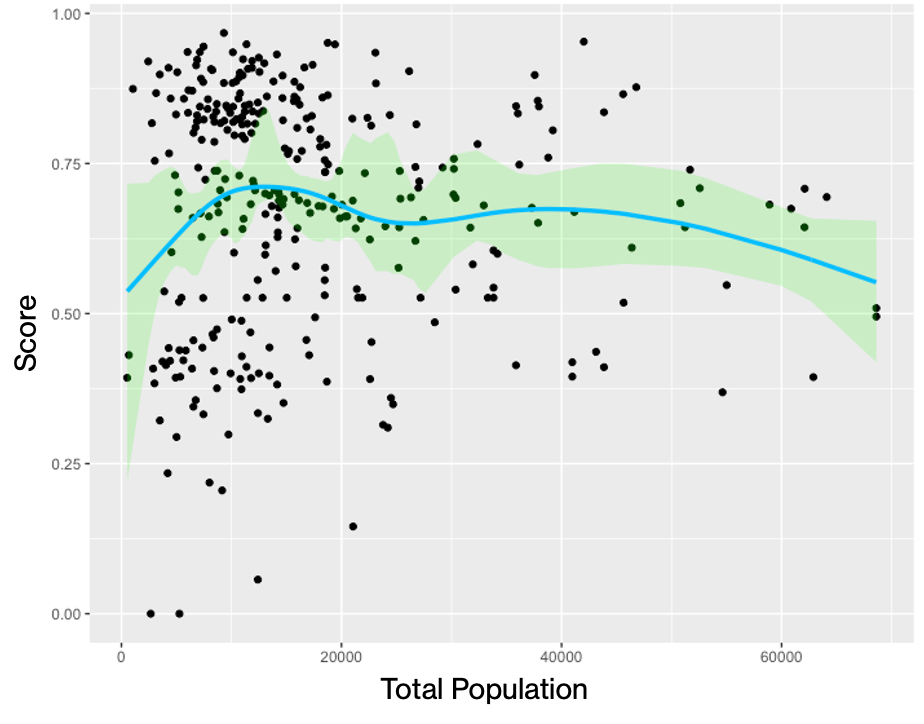}
    \caption{ADRF for total population in a buffer region around the trap.}
    \label{fig:6a}
    \end{subfigure}
    \hfill
    \begin{subfigure}{0.32\textwidth}
         \centering
    \includegraphics[width=\linewidth]{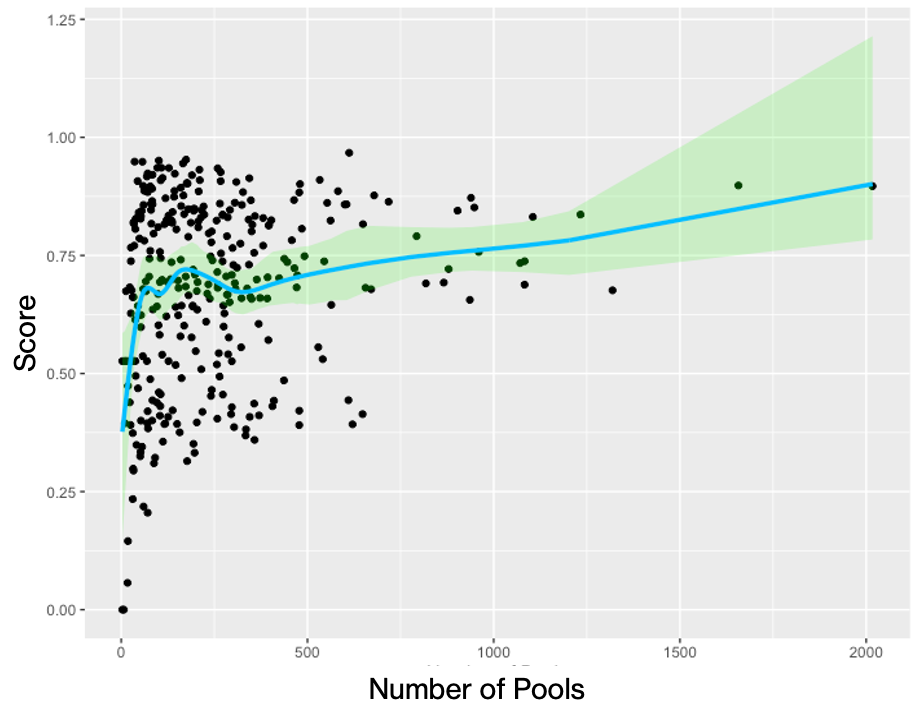}
    \caption{ADRF for the number of pools collected from the trap.}
    \label{fig:6b}
    \end{subfigure}
    \hfill
    \begin{subfigure}{0.32\textwidth}
    \centering
    \includegraphics[width=\linewidth]{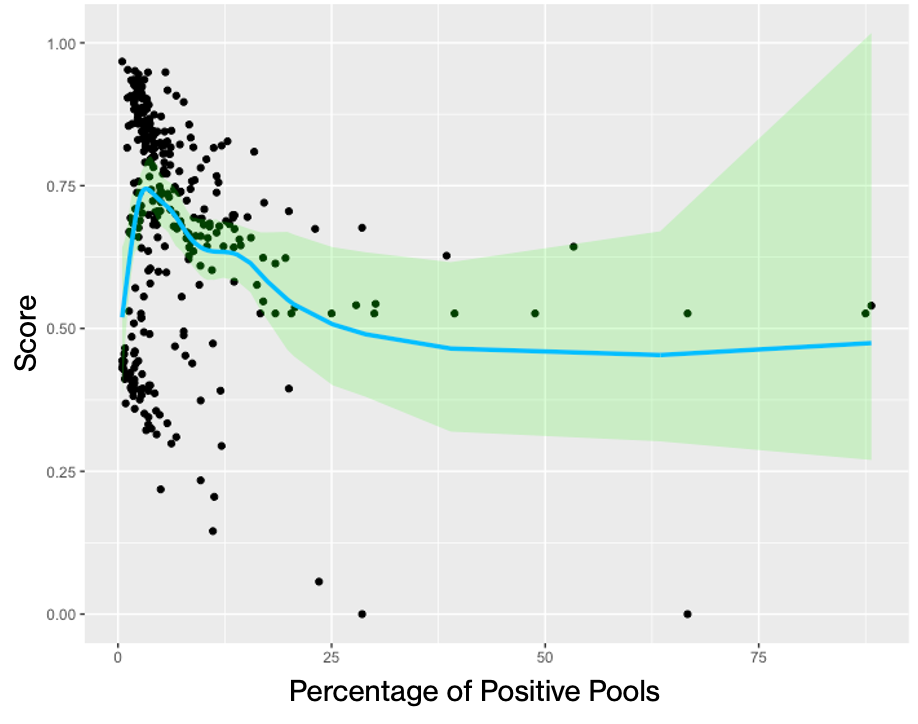}
    \caption{ADRF for the number of positive pools collected from the trap.}
    \label{fig:6c}
\end{subfigure}
    \caption{Estimate of the ADRF for a variable's causal effect on the score with 95\% pointwise standard errors (denoted in green). The standard errors are estimated by bootstrapping the entire estimation process.}
    \label{fig:ADRF}
\end{figure}

Figure \ref{fig:ADRF} shows the estimates for the average dose-response functions (ADRFs) calculated from Phase 3 of our method (as described in Section \ref{sec:Phase 3}). Interpretation of these ADRFs is straightforward: a horizontal line suggests no causal effect, while an ascending or descending line indicates a positive or negative causal effect, respectively. In Figure \ref{fig:ADRF}, we only show ADRFs of the three variables that exhibit a significant causal effect. The first is the total population in a 1500m buffer region around the traps. Upon examining the ADRF (figure \ref{fig:6a}), we find a positive causal effect of the variable on the score until approximately a population of 10,000, after which the curve levels off. This implies that traps in locations with higher populations contribute to higher scores up to a certain population threshold, indicating that traps placed in such areas would be more effective at predicting human cases than traps placed in areas of low population.
This result does not necessarily mean that the mosquitoes in populated areas have more tendency to carry WNV, but it mainly points out that mosquito traps in those areas are more useful as an indicator for human cases. 

The second variable demonstrating a significant causal effect is the number of pools collected from a trap. This variable is highly correlated with the mosquito population near the trap, as a larger number of pools often results from too many mosquitoes collected.
Figure \ref{fig:6b} shows a positive causal effect of the number of pools collected from the trap on the score. 
This finding is intuitive, as areas with higher mosquito populations (resulting in more pools per trap night) pose a potentially greater risk for WNV transmission. However, this relationship could also be the result of traps deployed for more trap nights providing more data for analysis, which could increase the stability of the sensitivity and specificity estimates. It would be worthwhile to compare traps with similar numbers of trap nights to determine if the effect of the number of pools remains; however, this would greatly reduce the sample size in this dataset and result in limited statistical power. 

Figure \ref{fig:6c} depicts the causal effect of the percentage of pools that tested positive for WNV on trap scores. 
This figure illustrates an initial positive effect, succeeded by a diminishing impact that eventually levels off. Given that the score is a weighted average of sensitivity and specificity, a very low or high percentage of positive pools can both adversely influence the score. This is because the lack of variability in pool results often leads to a lack of variability in prediction, resulting in either a high sensitivity and low specificity (always predicting cases) or a low sensitivity and high specificity (never predicting cases).
Thus, a location receiving a balanced mix of positive and negative test results is expected to be the most effective, as shown in the last ADRF plot, because such a location would be less likely to produce high numbers of false positives or false negatives. While this result does not help identify entirely new locations, 
it does help identify currently utilized trap locations that are less effective. Redirecting resources from traps that constantly yield only negative or positive results to locations with more variability in results can improve surveillance efforts by increasing the information gained from each trap night. All the other variables considered in the analysis showed no causal effect on the score.

\section{Additional application: Identifying features affecting specificity}

As mentioned in the introduction, one major contribution of our work lies in the development of a flexible three-phase modelling framework that can be adapted to various problems. We have demonstrated the ability of our model to identify optimal trap locations to enhance efficiency. However, beyond its primary application in trap placement optimization, our method offers broader potential for enhancing WNV surveillance practices.  In this section, we explore an additional application of our method, 
showcasing how it can be utilised to gain further insights and identify potential shortcomings in WNV surveillance within the region. 

\begin{figure}[ht]
    \centering
    \begin{subfigure}{0.48\textwidth}
        \centering
    \includegraphics[width=\linewidth]{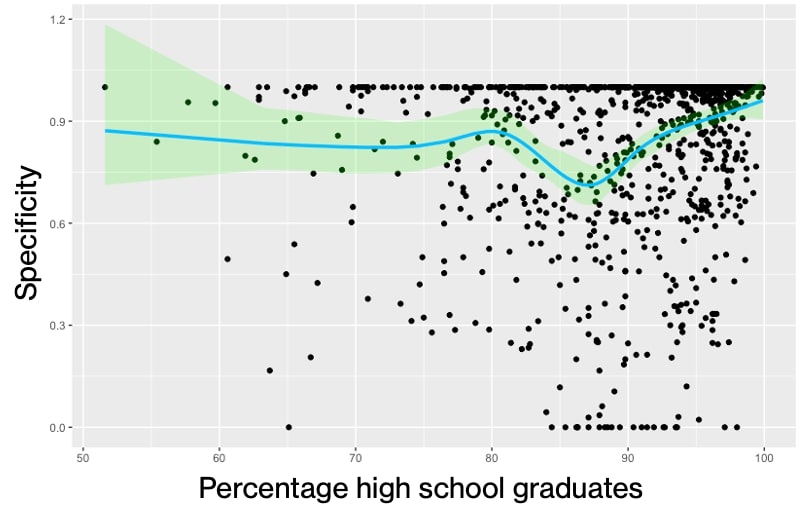}
    \caption{ADRF for the percentage of high school graduates in a buffer region around the trap.}
    \label{fig:7a}
    \end{subfigure}
    \hfill
    \begin{subfigure}{0.48\textwidth}
         \centering
    \includegraphics[width=\linewidth]{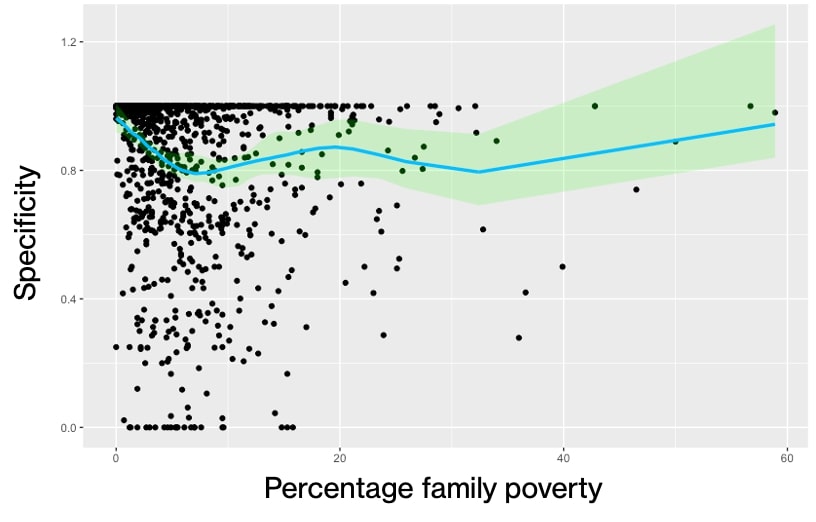}
    \caption{ADRF for the percentage of family poverty in a buffer region around the trap.}
    \label{fig:7b}
    \end{subfigure}
    \caption{Estimate of the ADRF for a variable's causal effect on the score with 95\% pointwise standard errors (denoted in green). The standard errors are estimated by bootstrapping the entire estimation process.}
    \label{fig:ADRFspec}
\end{figure}

Up to this point, we have solely utilized the 326 traps in $T^*$ to ascertain the factors contributing to the identification of the best trap locations. The selection of $T^*$ was mainly based on whether we have the data to calculate the sensitivity values of the traps, as the score we defined is a weighted mean of the sensitivity and specificity values. However, to obtain further insights from the trap data, we now expand our analysis to incorporate all available traps (1026 traps $\in T$) by revising the score definition for an individual trap $t \in T$ as:
$$\hbox{score}'_t = \hbox{AvgSpec}_t.$$
With this updated score definition, we reapply Phase 3 which involves conducting a causal analysis as outlined in Section \ref{sec:Phase 3}, to identify the variables influencing the specificity (the updated score) of a trap. Essentially, we aim to determine which features lead to a trap falsely predicting the incidence of a human case of WNV. In our analysis, two features emerge as particularly interesting for their causal effect on specificity: poverty levels and the proportion of high school graduates in the vicinity of the trap. The ADRFs for these features plotted against specificity are depicted in Figure \ref{fig:ADRFspec}.

Our analysis reveals notable trends regarding the impact of socioeconomic factors on trap specificity. Specifically, we observe an increasing trend in the ADRF for the percentage of high school graduates (figure \ref{fig:7a}) and a decreasing trend for the percentage of family poverty (figure \ref{fig:7b}). This suggests that in regions with high poverty levels and low levels of education, traps tend to predict the incidence of cases even in the absence of reported cases. This finding is particularly significant as it may indicate potential underreporting of WNV cases in these areas. Underreporting of WNV cases is a pressing issue, particularly in communities with limited access to healthcare services due to economic constraints and geographical barriers. Our analysis of the trap data not only provides evidence supporting this concern but also helps pinpoint specific locations where this problem is particularly pronounced.

\comment{
Package - causaldrf
Contains doubly robust methods for estimating ADRF
Using a GAM estimate - This estimates the ADRF using a method similar to that described in Hirano and Imbens (2004) but with spline basis terms in the outcome model.
In particular, treatment modelled on covariates as 
\begin{equation}
    T_i|X_i \sim N(X_i^T\beta, \sigma^2)
\end{equation}

This generates the general propensity scores
\begin{equation*}
    \hat{R_i} = \frac{1}{\sqrt{2\pi\hat{\sigma}^2}}\exp\left(-\frac{(t-X_i^T\hat\beta^2)}{2\hat{\sigma}^2}\right)
\end{equation*}
These GPS estimates become predictors in the outcome model. 
The outcome is modelled as a function of $T_i$ and $\hat R_i$ parametrically using GAM.

\[\Expect{Y_i|T_i,X_i}  \sim 1 + T_i +\hat R_i + \hat R_iT_i\]
}

\section{Discussion and conclusion}
This study presents a method to evaluate the monitoring locations used for environmental surveillance, in the context of identifying optimal mosquito trap locations for prediction of human WNV cases. The problem of where to collect environmental surveillance samples is long-standing in several fields. Physical samples often cannot be collected from a comprehensive grid across the surveilled area due to logistical and cost limitations. Our method can offer two solutions for this particular problem. First, the score computation from Phase 2 can identify low-performing surveillance locations, those with low scores. Second, the results from Phase 3 can guide the process of finding new surveillance locations. It is hoped that this latter solution will allow low-resourced areas to prioritize sampling based on the results from higher-resourced areas. 
Furthermore, our model offers the benefit of generalizability. For example, it allows us to provide additional insights into the WNV surveillance process, by defining the score differently and considering diverse features to assess their causal impact on the new score, similar to the analysis done in Section 5. Those new insights regarding the trap data can improve our understanding of WNV surveillance in ways previously unexplored. 

While there exist previous studies which have explored trap placement, they have typically been constrained by limited scope and resources. For example, in \citet{crepeau2013effects} researchers studied trap performance for Asian tiger mosquitoes by placing traps in locations with varying light, temperature, and humidity levels. Such studies on trap placement are typically limited to a small number of features due to their experimental nature which can be costly. In contrast, our approach allows for the inclusion of any variable considered important for the analysis. Additionally, there have been studies on researching the optimal trap placement in different contexts. \citet{smith2023mgsurve} examined optimal trap placement for genetic surveillance of mosquito populations, which differs from the context of our investigation. 

Our analysis reveals several key findings: Firstly, regions with higher population densities are recommended for increased trap placement. Additionally, areas having higher mosquito densities (indicated by traps capturing a greater number of mosquito pools), and those receiving a balanced mix of positive and negative test results from traps are identified as particularly effective trap locations. Interestingly, our analysis found that the landscape of trap locations does not play a role in their performance. This corroborates results from \citet{karki2016effect} which found that for gravid traps, the landscape features make little contribution to the variability in mosquito abundance. 

Our study also provides evidence of possible underreporting of WNV cases in areas characterized by higher levels of poverty or lower levels of education. This highlights the need for targeted surveillance efforts in socioeconomically disadvantaged communities to mitigate the risk of disease transmission and improve public health outcomes.

An important caveat of our approach is that we need to make sure the causal analysis is done correctly using the directed acyclic graphs (DAGs). Confounding variables have the potential to significantly impact results, introducing spurious associational relationships that may ultimately lead to erroneous conclusions if treated causally. Thus, careful attention to confounders is crucial to ensure the validity and reliability of the causal inferences drawn from the analysis. Additionally, it should be noted that this analysis is based on the mosquito testing data provided by vector control, which may not necessarily represent all mosquitoes collected in the trap. This could result in poor predictive ability for these traps. However, the majority of the traps are within the mosquito abatement districts in the Chicago area, which have a policy of testing all Culex spp. mosquitoes collected (Personal communication).

In conclusion, we introduce a novel modelling framework to find the best location to place traps for efficient WNV surveillance. The fundamental benefit of our method is that it is versatile, adaptable to various scenarios, and extendable to tackle challenges beyond trap placement optimization. 

\section{Acknowledgement}
This publication was supported by Cooperative Agreement Number U01CK000651 from the Centers for Disease Control and Prevention. Its contents are solely the responsibility of the authors and do not necessarily represent the official views of the Centers for Disease Control and Prevention or the Department of Health and Human Services.

{
\bibliography{references.bib}
\bibliographystyle{apalike}
}

\end{document}